\documentclass[twocolumn]{aastex6}
\usepackage{float}
\usepackage{graphicx}
\usepackage{amsmath}
\usepackage{natbib}
\usepackage{color}
\usepackage{verbatim}
\usepackage[utf8]{inputenc}
\usepackage{mathtools}
\usepackage{amssymb}
\usepackage{textcomp}
\usepackage{totcount}

\newtotcounter{citenum}
\def\oldbibitem{} \let\oldbibitem=\bibitem
\def\bibitem{\stepcounter{citenum}\oldbibitem}

\DeclareUnicodeCharacter{2212}{\textminus}

\newcommand{\lcdm}{$\Lambda$\textsc{CDM}}

\slugcomment{The Astrophysical Journal 863, 152 (2018) \hspace{0.5cm} Published 20 August 2018}
\shortauthors{Smercina \textit{et al.}}

\begin{document}
\title{A Lonely Giant: The Sparse Satellite Population of M94 \\
Challenges Galaxy Formation
}

\author{%
Adam Smercina\altaffilmark{1},
Eric F. Bell\altaffilmark{1}, 
Paul A. Price\altaffilmark{2}, 
Richard D'Souza\altaffilmark{1,3},
Colin T. Slater\altaffilmark{4},\\
Jeremy Bailin\altaffilmark{5},
Antonela Monachesi\altaffilmark{6,7},
David Nidever\altaffilmark{8,9},
}

\altaffiltext{1}{Department of Astronomy, University of Michigan, 1085 S. University Avenue, Ann Arbor, MI 48109-1107, USA; asmerci@umich.edu}
\altaffiltext{2}{Department of Astrophysical Sciences, Princeton University, Princeton, NJ 08544, USA}
\altaffiltext{3}{Vatican Observatory, Specola Vaticana, V-00120, Vatican City State}
\altaffiltext{4}{Astronomy Department, University of Washington, Box 351580, Seattle, WA 98195-1580, USA}
\altaffiltext{5}{Department of Physics and Astronomy, University of Alabama, Box 870324, Tuscaloosa, AL 35487-0324, USA}
\altaffiltext{6}{Departamento de F\'isica y Astronom\'ia, Universidad de La Serena, Av. Juan Cisternas 1200 N, La Serena, Chile}
\altaffiltext{7}{Instituto de Investigaci\'on Multidisciplinar en Ciencia y Tecnolog\'ia, Universidad de La Serena, Ra\'ul Bitr\'an 1305, La Serena, Chile}
\altaffiltext{8}{Department of Physics, Montana State University, P.O. Box 173840, Bozeman, MT 59717-3840}
\altaffiltext{9}{National Optical Astronomy Observatory, 950 North Cherry Ave, Tucson, AZ 85719}

\begin{abstract}
The dwarf satellites of `giant' Milky Way (MW)-mass galaxies are our primary probes of low-mass dark matter halos. The number and velocities of the satellite galaxies of the MW and M31 initially puzzled galaxy formation theorists, but are now reproduced well by many models. Yet, are the MW's and M31's satellites representative? Were galaxy formation models `overfit'? These questions motivate deep searches for satellite galaxies outside the Local Group. We present a deep survey of the `classical' satellites ($M_{\star}$$\geqslant$4$\times$10$^5 M_{\odot}$) of the MW-mass galaxy M94 out to 150\,kpc projected distance. We find \textit{only two} satellites, each with $M_{\star}{\sim}10^6 M_{\odot}$, compared with 6--12 such satellites in the four other MW-mass systems with comparable data (MW, M31, M81, M101). Using a `standard' prescription for occupying dark matter halos (taken from the fully hydrodynamical EAGLE simulation) with galaxies, we find that such a sparse satellite population occurs in $<$\,0.2\% of MW-mass systems --- a $<$\,1\% probability among a sample of five (known systems + M94). In order to produce an M94-like system more frequently we make satellite galaxy formation much more stochastic than is currently predicted by dramatically increasing the slope and scatter of the stellar mass--halo mass (SMHM) relation. Surprisingly, the SMHM relation must be altered even for halos masses up to 10$^{11}M_{\odot}$\ --- significantly above the mass scales predicted to have increased scatter from current hydrodynamical models. The sparse satellite population of this `lonely giant' thus advocates for an important modification to ideas of how the satellites around MW-mass galaxies form.
\end{abstract}

\section{Introduction}
\label{sec:intro}
While the $\Lambda$-Cold Dark Matter (\lcdm) paradigm successfully explains the large-scale properties of the Universe, many of its predictions on small scales appear to be in tension with the number and properties of dwarf galaxies \citep[see the review by][]{bullock2017}. While the Milky Way (MW) hosts $\sim$10 `classical' dwarf satellite galaxies with velocity scales of $\gtrsim$10\,km\,s$^{-1}$\ (see \citealt{mcconnachie2012}), dramatically more DM halos were predicted to exist at those scales --- the `Missing Satellites Problem' (MSP; \citealt{klypin1999,moore1999}). Later work appeared to sharpen the problem by suggesting that the velocities of the few satellites the MW does have are substantially lower than the velocity scales of the most massive predicted dark matter (DM) halos --- the `Too Big to Fail' problem (TBTF; \citealt{boylan-kolchin2011}). 

While it is possible that these observations may signal the need for an important modification to \lcdm, it is widely accepted that improved galaxy formation physics is the likely resolution to these problems. Feedback from supernovae is predicted to dramatically suppress the number of stars in even relatively massive DM halos ($M_h \sim 10^{10} M_{\odot}$) \citep{maccio2010,font2011}. Furthermore, supernovae-driven outflows can drag DM to larger radii and could reduce the central velocities of these halos to observed values \citep{brooks2013,wetzel2016}. Recently, it has been suggested that tidal disruption of satellites and their subhalos around massive galaxies like the Milky Way reduces the number of predicted satellites further \citep[e.g.,][]{garrison-kimmel2017b}. These models, along with more a more complete census of MW satellites from surveys such as SDSS and DES, have been so successful that it has been argued `there is no missing satellites problem' \citep{kim2017}.

Nearly all of our understanding of the small-scale challenges to \lcdm\ has been based on observations of satellites in the Local Group (LG), prompting important questions:
\begin{enumerate}
\item Are the LG's satellites representative?
\item Are galaxy formation models in turn representative, or have they instead been over-tailored to fit the LG's satellites?
\end{enumerate}

These questions motivate a deep census of the satellites of `giant' MW-mass galaxies and low-mass field dwarf galaxies. A variety of approaches are being taken: deep spectroscopic surveys \citep[e.g., ][]{spencer2014,geha2017}, field H\,\textsc{i} observations \citep{papastergis2015,yaryura2016}, large-area integrated light surveys around nearby galaxies \citep[e.g., ][]{chiboucas2009,muller2015,danieli2017}, and narrower and deeper surveys which allow satellites to be resolved into stars \citep[e.g., ][]{toloba2016,carlin2016,crnojevic2016,smercina2017}. Discovering and confirming a complete sample of even `classical' satellites around nearby galaxies requires a formidable combination of depth and area. Such a sample exists for only four MW-like systems: the MW and M31 \citep{mcconnachie2012}, M81 \citep{karachentsev&kudrya2014}, and M101 \citep{danieli2017}.

In this paper we present the discovery of two low-mass satellites of the nearby MW-mass galaxy M94 (NGC 4736; $M_{\star} \simeq$\ 4$\times$10$^{10} M_{\odot}$, \citealt{karachentsev2013}; $D = 4.2$\,Mpc, \citealt{radburn-smith2011}), detected in a deep Hyper Suprime-Cam (HSC) survey with an effective radius of 150\,kpc. Rather than discovering the $\sim$10 `classical' satellites which were expected by scaling the other MW-mass systems, \textit{we discovered only two}, both with stellar masses only $\lesssim$\,10$^6 M_{\odot}$\ --- a satellite population completely unlike any other known galaxy. We further show that, more than being unexpected, M94's satellite population cannot be explained using `standard' galaxy formation models --- directly advocating for significant modifications to the physics of low-mass halo occupation.

\section{Observations}
\label{sec:obs}
\begin{figure*}[t]
\centering
\leavevmode
\includegraphics[width={0.76\linewidth}]{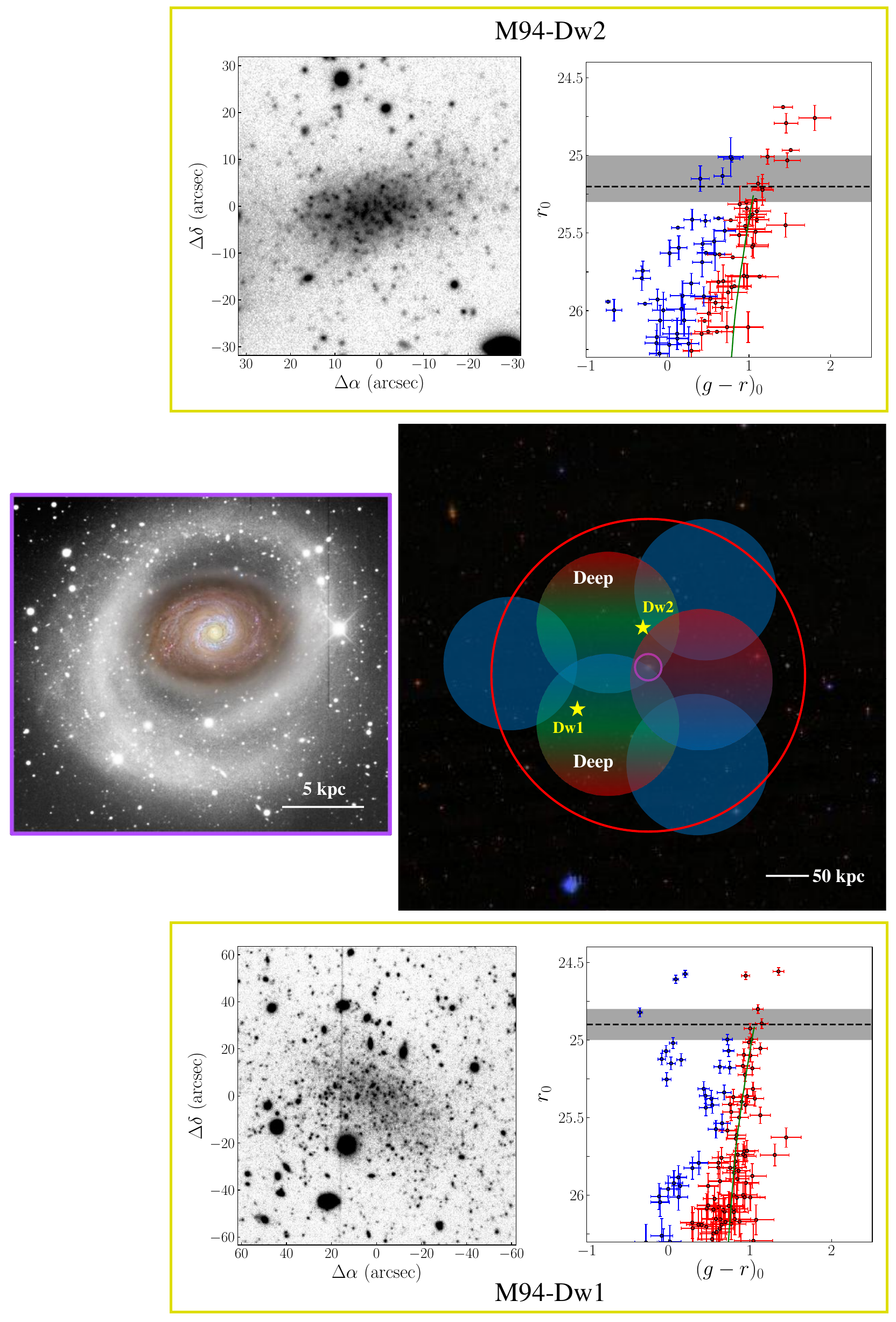} 
\caption{\uline{Right panel}: A $\sim$5$\times$5 $\square$\textdegree\ SDSS image centered on M94 (magenta circle). The colored circles show the six-pointing HSC survey footprint, while the red circle shows a circle of the same area with 150\,kpc `effective' radius. Blue denotes pointings observed in $g$-band, green in $r$-band, and red in $i$-band. The two deep pointings are labeled. The positions of Dw1 and Dw2 are shown as yellow stars. \uline{Bottom panel}: Deep $r$-band image of Dw1, accompanied by a CMD of detected stars in the dwarf. Red points represent RGB stars and blue points represent candidate core Helium-burning stars. The dashed line and gray region show the best-fit TRGB with uncertainty, while the green curve is the best-fit isochrone at that distance. \uline{Top panel}: Imaging and CMD for Dw2, following the same schema as for Dw1. \uline{Left panel}: Deep image of M94, taken from \cite{trujillo2009}.}
\label{fig:field}
\end{figure*}

Our HSC survey, carried out through the NOAO Gemini--Subaru exchange program (PI: Smercina, NOAO 2017A-0312), consists of six HSC pointings in $g$-band (for satellite discovery), three in $i$, and two in $r$ (for stellar halo characterization). The deepest two fields were observed for $\sim$7200s per filter in $gri$. The remaining four fields were observed for 1200s per filter. The average seeing FWHM in $g$-band is $\sim$0\farcs8 for all fields. The $g$-band survey footprint is symmetric around M94 and has an area equal to a 150\,kpc radius circular region, giving an `effective' radius of 150\,kpc (see Fig. \ref{fig:field}).

The data were reduced as described in \cite{smercina2017}, using the updated HSC pipeline \citep{bosch2018}. Data were calibrated using the Pan-STARRS1 survey \citep{magnier2013}, but all magnitudes are in the SDSS photometric system and have been corrected for foreground Galactic extinction using the updated \citet{schlafly2011} corrections. 

The satellite-focused part of the survey consists of six fields with moderately deep to very deep $g$-band data. These six fields were visually inspected for low surface brightness (SB) candidate dwarf galaxies with resolved or semi-resolved stars with luminosities and half-light radii similar to Local Group satellites following \cite{chiboucas2009} --- only two were found. One of these dwarfs, M94-Dw1, was detected as a dwarf galaxy candidate (dw1255+40) in the integrated light survey of \cite{muller2017}. The locations and properties of these dwarfs are given in Figure \ref{fig:field} and Table \ref{tab:dwarfs}.

\subsection{Completeness}
\label{sec:artificial}
In order to understand our satellite detection completeness limits we conducted $\sim$500 individual artificial galaxy tests in the $g$-band data. Satellites were inserted with a range of projected distances to M94 of $15<D/kpc<150$. Tests were done by eye on isolated images to eliminate subconscious cross-referencing with known areas in the field, and with multiple authors to produce independently verifiable results. Dwarf galaxies were simulated by sampling random stars from a 12\,Gyr isochrone and injecting them into the survey images. These artificial galaxies were created from single stellar populations and thus resemble quiescent satellites in the LG. The positions of stars were drawn randomly from independent Gaussian profiles in \textsc{X} and \textsc{Y}, with randomly-generated galaxy centers, elliptical half-light radii ranging from 0.1--1\,kpc, and corresponding ellipticities $\epsilon \leqslant 0.9$. Position angles were also chosen at random, ranging from 0\textdegree--360\textdegree. Additionally, in a given test there was a 30\% chance of not injecting an artificial galaxy. 

\begin{figure*}[t]
\centering
\leavevmode
\includegraphics[width={0.95\linewidth}]{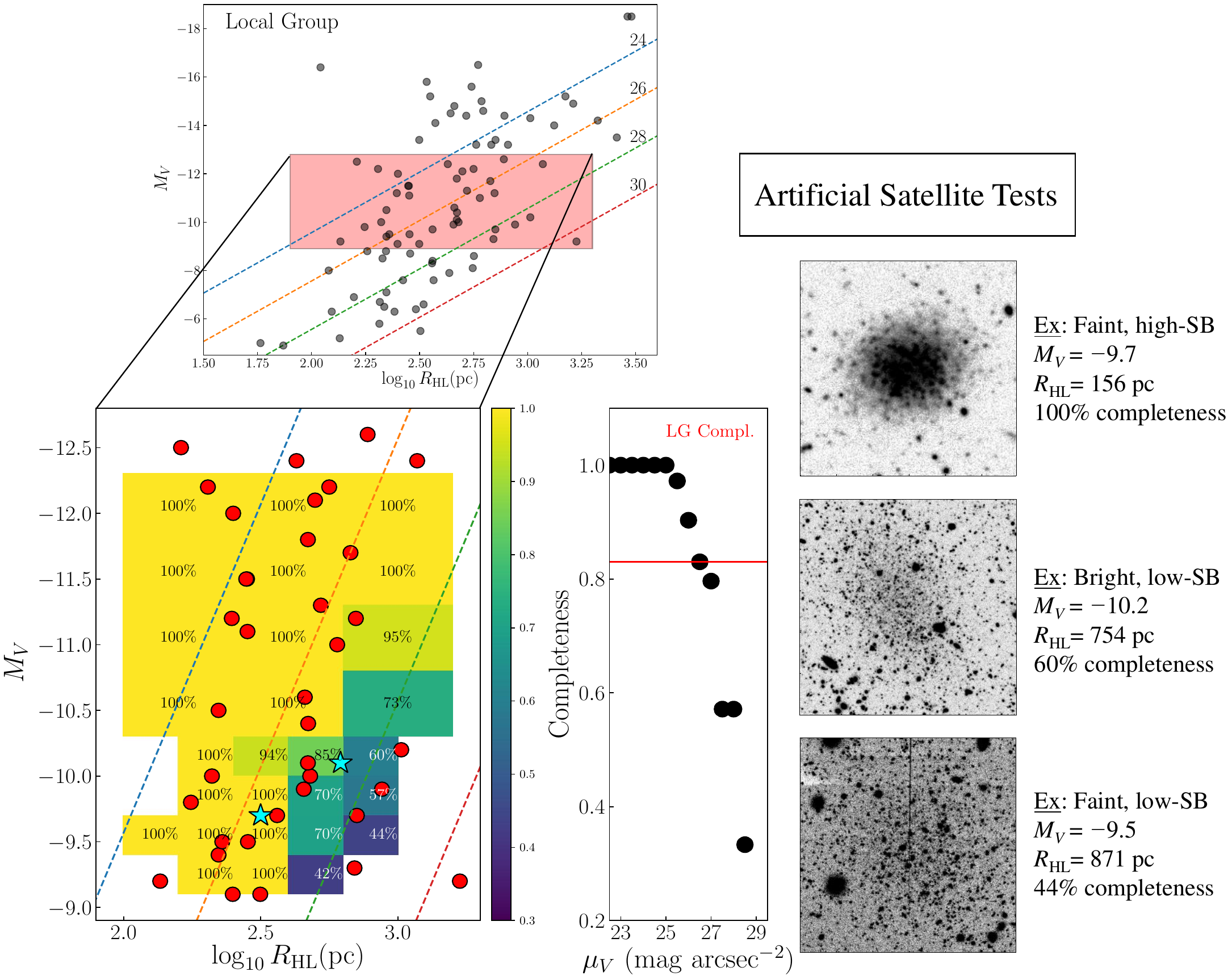} 
\caption{Results of our artificial satellite tests. \uline{Top left}: Size--luminosity relation for \textit{all} satellites of the Local Group in \cite{mcconnachie2012}. Lines of constant SB are shown at 24 (blue), 26 (orange), 28 (green), 30 (red) $M\,\square\arcsec^{-2}$. The red patch denotes the approximate region probed by our artificial satellite tests. \uline{Bottom left}: Recovery completeness map for injected artificial satellites in size--luminosity space. The red circles are LG satellites. The cyan stars represent Dw1 and Dw2. \textit{Right panel}: Completeness as a function of SB for artificial satellites. The red line shows our 85\% detection completeness for LG satellites in the range $-9.1 > M_{V} > -10.3$. \uline{Right}: Selected examples of detected artificial dwarfs in three different luminosity/SB regimes.}
\label{fig:fake-sat}
\end{figure*}

Most effort was spent sampling galaxies with luminosities $-9.1 > M_{V} > -10.3$, spanning the range of the two candidates --- constituting 300 tests. We conducted an additional 140 tests in the luminosity range $-10.3 > M_{V} > -12.3$. As further explained in \S\,\ref{sec:halo-occ}, we account for up to 1\,Mpc uncertainty in line-of-sight (LOS) distance between satellites and the central galaxy (M94 in this case). LOS distance variance only significantly affects the luminosity of satellites at the distance of M94 when it exceeds 0.4 Mpc ($\sim$0.2\,mag). From simulations (see \S\,\ref{sec:halo-occ}), the distribution of satellite LOS distances from MW-mass galaxies is well-fit by a Lorentzian distribution --- the large-distance wings stemming from the two-point correlation function of galaxies. We estimate that $\sim$10\% of apparent satellites around MW-mass galaxies likely have $\Delta\,d = \pm$0.4--1\,Mpc. To explore this effect on our test results, 30 additional tests were conducted, focused on satellites in the $-9.1 > M_{V} > -10.3$\ luminosity range ($\sim$10\% of our original 300 tests in that luminosity range), and placed at varying distances, drawn from a Lorentzian distribution with $\Delta\,d = \pm$0.4--1\,Mpc.

Figure \ref{fig:fake-sat} shows the results of our tests, with three example artificial satellites. Overall completeness was high, with $>$70\% of all injected satellites recovered. The rate of false detection was \textit{very} low (2\%) and only occurred in the very lowest-surface brightness ($\mu_{V} \lesssim 28\ M\,\square\arcsec^{-2}$) cases. Completeness is a relatively smooth function of SB (middle panel), with $\sim$80\% completeness corresponding to a SB of $\mu_{V} = 27\ M\,\square\arcsec^{-2}$. Overall completeness is $\sim$10\% higher on average in the two deeper fields, but otherwise there is little dependence on field location, even when binning by SB. The lower left panel shows the completeness binned as a function of luminosity and half-light radius, along with the LG satellites with luminosities approximately within our test range. Applied to the properties of LG satellites, we estimate an average completeness of 85\% in the range $-9.1 > M_{V} > -10.3$. Our satellite tests in the range $-10.3 > M_{V} > -12.3$\ yield a very high 97\% average completeness, and is $\sim$100\% when applied to LG satellites. The effect of LOS distance variance was negligible, following our 30 additional tests. Average completeness remained the same, likely owing to the competing effects of lower completeness at farther LOS distances and proportionally higher completeness at closer distances. 

Scaling from the mass-to-light ratios of LG satellites \citep{mcconnachie2012}, our 85\% completeness limit of $M_{V} = -9.1$\ corresponds to $M_{\star}{\sim}$4$\times$10$^{5} M_{\odot}$. Consequently, M94, a MW-mass galaxy, very likely hosts only two satellite galaxies with projected radii $<$\,150\,kpc and $M_{\star} \gtrsim 4$$\times$$10^5 M_{\odot}$\ --- a satellite population unlike any other known galaxy of its kind. 

\begin{deluxetable}{lll}
\tablecaption{\textnormal{Dwarf Parameters}\label{tab:dwarfs}}
\tablecolumns{3}
\setlength{\tabcolsep}{8pt}
\setlength{\extrarowheight}{0.7pt}
\tabletypesize{\small}
\tablehead{%
\colhead{Parameter} &
\colhead{M94-Dw1$^{a}$} & 
\colhead{M94-Dw2} \\
}
\startdata
$\alpha$ (J2000) & $12^{h} 55^{m}02\fs 49$ & $12^{h} 51^{m}04\fs 4$ \\
$\delta$ (J2000) & $40^{\circ} 35\arcmin 21\farcs 9$ & $41^{\circ} 38\arcmin 09\farcs 9$ \\
$D_{\textrm{TRGB}}$ & 4.1\,$\pm$\,0.2 Mpc & $4.7^{+0.2}_{-0.4}$\ Mpc \\
$M_{V}~\!\!^{b}$ & \textminus10.1\,$\pm$\,0.1 & \textminus9.7$^{-0.1}_{+0.2}$ \\
$r_{h}$ & 618,$\pm$\,90 pc & 316\,$\pm$\,40 pc \\
$\mu_{V,eff}~\!\!^{c}$ & 27.4 $M\,\square$\arcsec$^{-1}$ & 26.4 $M\,\square$\arcsec$^{-1}$ \\
$M_{\star}~\!\!^{d}$ & 9.7$\times$10$^5 M_{\odot}$ & 6.7$\times$10$^5 M_{\odot}$ \\
$[$Fe/H$]~\!\!^{e}$ & \textminus2.1\,$\pm$\,0.1 & \textminus2.1\,$\pm$\,0.1 \\
\enddata
\tablecomments{$^{a}$~Also dw1255+40 in \cite{muller2017}. $^{b}$~Profile fitting, assuming $D_{\textrm{TRGB}}$. $^{c}$~Effective $V$-band surface brightness within the half-light radius. $^{d}$~Comparing to dwarf irregulars of similar luminosity in the Local Group \citep{mcconnachie2012}. $^{e}$~Metallicity of best-fit isochrone, assuming $[\alpha$/Fe] = 0.25.} 
\end{deluxetable}

\section{Satellite Properties}
\label{sec:dwarfs}
The two dwarfs were detected in the two fields with $gri$\ imaging, allowing for analysis of their stellar populations. In Figure \ref{fig:field} we show the $r$-band images and color--magnitude diagrams (CMDs) of the dwarfs. Aperture photometry using successive elliptical apertures was used to construct brightness profiles and a total flux for each dwarf. The profiles were also used to determine half-light radii. $g$\ and $r$-band magnitudes were converted to $V$-band using the SDSS `Lupton 2005' photometric transformation\footnote{{\href{http://www.sdss.org/dr12/algorithms/sdssubvritransform}{http://www.sdss.org/dr12/algorithms/sdssubvritransform}}}.

Distances for the dwarfs were determined from the tip of the red giant branch (TRGB), estimated using a maximum-likelihood analysis following Appendix C of \cite{monachesi2016} and \cite{smercina2017}. We determined $r$-band completeness using artificial stars for the highly crowded regions of M94-Dw2. Dw1 and Dw2 are 4.1$\pm0.2$\,Mpc and 4.7$^{+0.2}_{-0.4}$\,Mpc away, both reasonably consistent with M94 group membership ($D_{\rm M94}{\sim}4.2$\,Mpc). We thus estimate absolute $V$-band magnitudes of \textminus10.1 and \textminus9.7, with 0.1--0.2\,mag uncertainties dominated by the TRGB distance. Projected distances from M94 are 69\,kpc for Dw1 and 38\,kpc for Dw2. 

Both Dw1 and Dw2 have stars bluer than the RGB with colors typical of young main sequence and intermediate-age core helium-burning stars \citep{radburn-smith2011}, indicating ongoing star formation. Furthermore, both dwarfs have irregular morphologies characteristic of star-forming galaxies at similar magnitudes \citep{carrillo2017}. While isolated dwarf galaxies are invariably star-forming \citep{geha2012}, the vast majority of LG satellites are quiescent \citep{slater&bell2014} --- assumed to be due to ram-pressure stripping during infall \citep[e.g.,][]{emerick2016,simpson2017}. Consequently, the star formation in M94's two satellites, with projected distances $<$\,100\,kpc, is puzzling. If these galaxies are shown to be significantly further from M94, it would make their star formation easier to understand, but would mean that M94 hosts even fewer satellites within its virial radius. Alternatively, this may indicate that M94 lacks the hot gas required to strip gas from satellites \citep{slater&bell2014}.

Given Dw1 and Dw2's $V$-band luminosities and a stellar $M/L_{V}{\sim}1$\ for similar star-forming dwarf galaxies in the LG (following \citealt{mcconnachie2012}), we estimate stellar masses of 9.7$\times$10$^{5} M_{\odot}$\ and 6.7$\times$10$^{5} M_{\odot}$.

Metallicities were determined by fitting PARSEC isochrone models \citep{bressan2012} to the $g-r$\ colors and $r$-band magnitudes, with a fixed 12 Gyr age and metallicities in the range $Z =$\ 0.0001--0.001. The best-fit isochrones, placed at the respective TRGB distances for each dwarf, each have metallicity $Z = 0.0002$, corresponding to an iron abundance of [Fe/H] $= -2.1$, assuming an [$\alpha$/Fe] $= 0.25$. This is consistent with the RGB-derived metallicities of similarly-massive star-forming dwarf galaxies in the LG (e.g., Sagittarius dIrr; \citealt{mcconnachie2012}).

\begin{figure*}[t]
\centering
\leavevmode
\includegraphics[width={0.95\linewidth}]{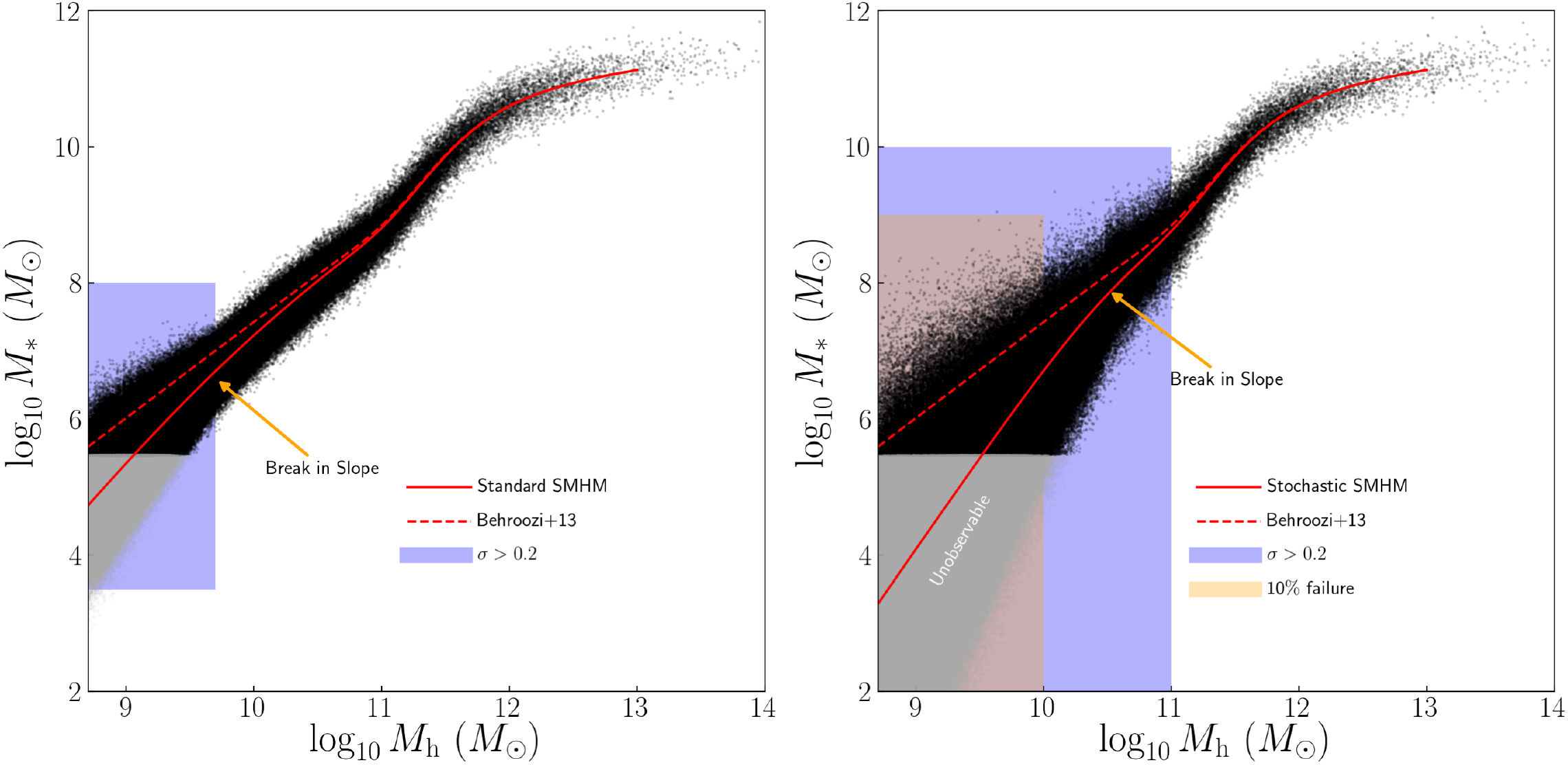} 
\caption{\uline{Left}: The SMHM relation for DM halos in EAGLE using `standard' halo occupation. The dashed red curve is taken from \cite{behroozi2013}. A standard 0.2\,dex log-normal scatter is assumed for $M_{\rm h,peak} > 5$ $\times$10$^9 M_{\odot}$. Below this mass, increased mass-dependent scatter and a steeper slope are adopted following \cite{munshi2017}. Gray points denote galaxies which are likely unobservable in our survey of M94. \uline{Right}: A radically altered SMHM relation, reflecting the stochastic halo occupation implied by M94's sparse satellite population. Increased, mass-dependent scatter is adopted for all halos with $M_{\rm h,peak} < 10^{11} M_{\odot}$. A significantly steeper slope is also assumed for halos with $M_{\rm h,peak} <$\ 3$\times$10$^{10} M_{\odot}$, along with a fixed 10\% rate of galaxy failure for $M_{\rm h,peak} < 10^{10} M_{\odot}$.} 
\label{fig:smhm}
\end{figure*}

\section{Implications for Galaxy Formation}
\label{sec:halo-occ}
As discussed in \S\,\ref{sec:intro}, the satellite populations of only four other MW-mass galaxies are known down to $<$\,10$^6 M_{\odot}$\ with good completeness: the MW, M31, M81, and M101 --- all central galaxies in low-density environments. Among these four, the average number of satellites with $M_{\star} >$\ 4$\times 10^5 M_{\odot}$, within a \textit{projected} 150\,kpc radius from the central, is $9 \pm 3$. Projected distances for MW and M31 satellites were determined using the derived physical LG coordinates of \cite{pawlowski2013}, and simulating 10,000 random LOS's from external reference positions. All four galaxies also host at least one satellite with $M_{\star} > 10^9 M_{\odot}$. Placed in context with these systems, the satellite population of M94, a MW-mass central galaxy in a low-density environment, is completely unexpected --- possessing only two `classical' satellites, with a most massive satellite of only $\sim$10$^6 M_{\odot}$. However, was such a system predictable in simulations? 

While a thorough theoretical analysis is beyond the scope of this paper, we explore the implications of our results for galaxy formation models using a simple halo occupation approach. For such an exercise, we require a simulation which a) provides a large diversity of accretion histories for MW-mass halos, b) resolves dark matter subhalos capable of hosting the satellites we are interested in ($M_{\rm h,peak} > 10^9\,M_{\odot}$), and c) can accurately account for subhalo disruption due to the potential of the central disk \citep[e.g.,][]{garrison-kimmel2017b}. The current generation of large-volume cosmological hydrodynamical simulations best meet these criteria. Here we use the dark matter subhalos of the large-volume, (${\sim}100\,{\rm Mpc}^{3}$) fully hydrodynamical version of the EAGLE simulation \citep{schaye2015}. We confirm the robustness of EAGLE's subhalo catalogs at low masses by comparing the average subahlo mass function to a higher-resolution simulation ($\sim 25\,{\rm Mpc}^{3}$) also made available by the EAGLE collaboration, finding that they converge for $M_{\rm h,peak} > 10^9\,M_{\odot}$. This is more than sufficient for our purposes, as most current models predict that cosmic reionization and stellar feedback should produce mostly `dark' halos below $M_{\rm h,peak} < 10^9 M_{\odot}$ \citep[e.g.,][]{sawala2015,ocvirk2016,munshi2017}. We choose not to directly use the stellar masses and properties of satellites from EAGLE (see e.g., \citealt{shao2018}) in this analysis, primarily because we need to explore the impact of varying prescriptions about how satellite galaxies populate dark matter halos.

\begin{figure*}[!ht]
\centering
\leavevmode
\includegraphics[width={0.95\linewidth}]{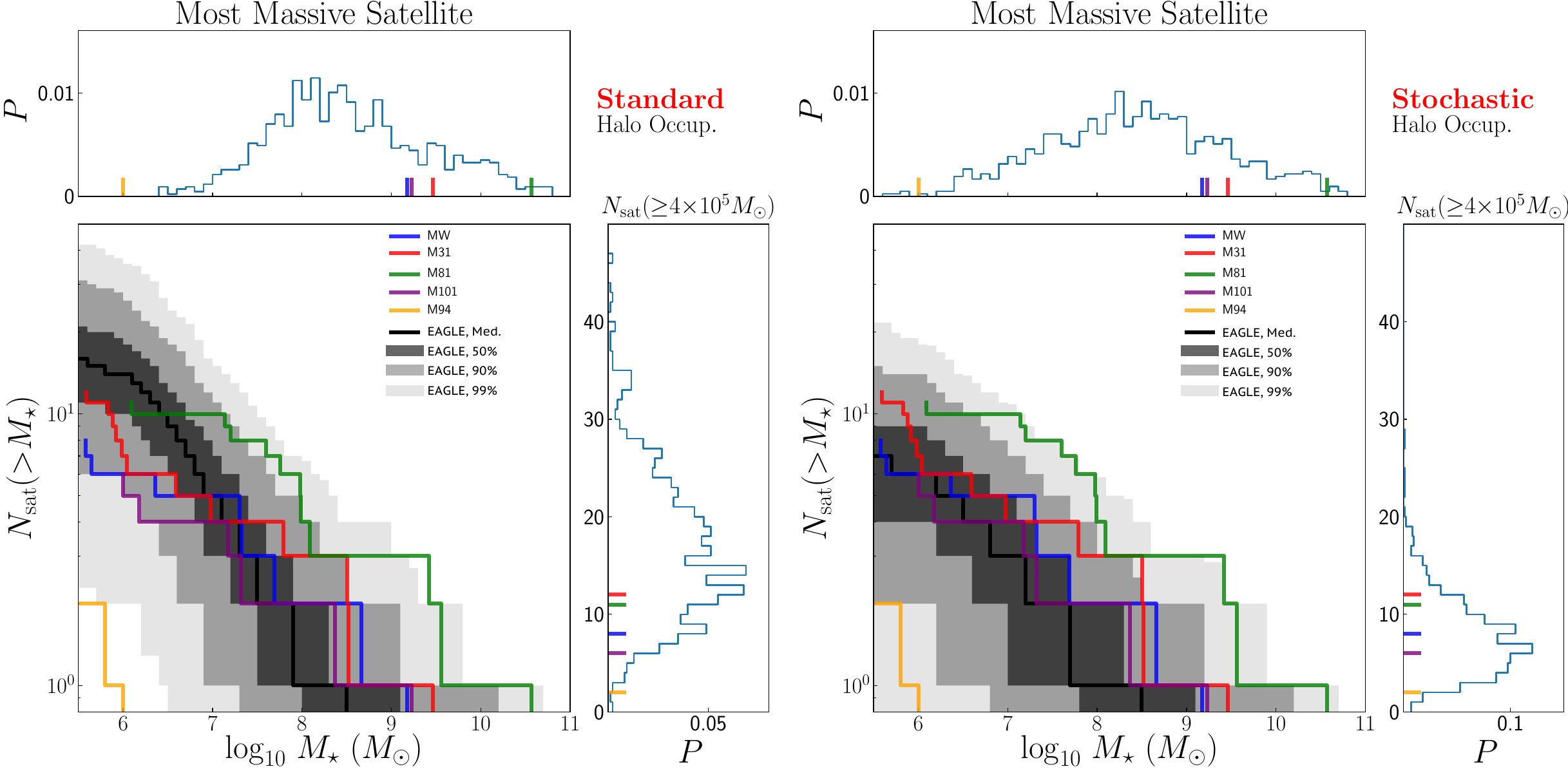} 
\caption{Satellite stellar mass functions and statistics for M94 and other nearby galaxies and EAGLE halos, assuming two halo occupation models. \uline{Left}: Satellite mass functions for nearby galaxies: M94 (orange), the MW (blue), M31 (red), M81 (green), and M101 (purple). Also shown are the median (black line) and 50\% (dark gray), 90\% (gray), and 99\% (light gray) confidence intervals for simulated satellite mass functions for MW-mass galaxies in EAGLE (completeness-corrected for $M_{*} < 10^6\ M_{\odot}$), assuming the `standard' halo occupation described in Figure \ref{fig:smhm}. `Standard' halo occupation produces M94-like systems $<$\,1\% of the time. \textit{Top panel}: Normalized histogram of the most massive satellite formed around each central EAGLE. Known galaxies are shown by vertical lines. \textit{Right panel}: Normalized histogram of the total number of $M_{\star} > 4$ $\times$10$^5 M_{\odot}$\ satellites for each central in EAGLE. Known galaxies are shown by horizontal lines. \uline{Right}: Same as the left panel, but assuming stochastic halo occupation. The shape of the mass function and subsequent distribution of the total number of satellites has changed dramatically, producing M94-like systems $>$4\% of the time.}
\label{fig:mfunc}
\end{figure*}

We assign galaxies to dark matter halos and subhalos using their `peak' (or infall) mass (denoted as $M_{\rm h,peak}$). Our halo occupation model, which applies equally for both central and satellite galaxies, follows the commonly-adopted \cite{behroozi2013} SMHM relation, with a fixed 0.2\,dex log-normal scatter, down to $M_{\rm h,peak}\,{\geqslant}$\,5$\times 10^{9} M_{\odot}$. The SMHM relation at low masses is very uncertain and an extrapolation of the \cite{behroozi2013} SMHM relation over-predicts the number of dwarf satellites of the Milky Way \citep[e.g.,][]{dooley2017}. Consequently, we adopt a somewhat steeper slope with increased mass-dependent scatter for $M_{\rm h,peak} <$\ 5$\times$10$^9 M_{\odot}$, following \cite{munshi2017}. Figure \ref{fig:smhm} (left panel) shows the adopted relationship between halo/subhalo mass and galaxy stellar mass for EAGLE dark matter halos.  

Next, we define `MW-mass galaxies' to be central halos with 6$\times$10$^{11} M_{\odot} \leqslant M_{\rm h,peak} \leqslant$\ 3$\times$10$^{12} M_{\odot}$, which host a galaxy with model-derived stellar mass of $M_{\star} \geqslant 4$ $\times 10^{10} M_{\odot}$\ --- a sample of 1,500 galaxies. As these halos are centrals, they automatically exclude halos in dense environments (cluster members or other satellites), but otherwise span a range of large-scale environments. In turn, we define `satellites' within a range of projected radii $15$\,kpc\,${<}\,D_{proj}\,{<}150$\,kpc from each EAGLE MW-mass central, and within 1\,Mpc in LOS (\textsc{Z}) distance --- a realistic observational constraint for satellites around nearby galaxies. 

Figure \ref{fig:mfunc} (left panel) shows the resulting satellite mass function for MW-mass galaxies in EAGLE, against known satellite mass functions within 150\,kpc projected distance from the central. The simulated satellite mass functions have been completeness-corrected to match our results for M94 (see \S\,\ref{sec:artificial}) --- 85\% for $4{\times}10^5\ M_{\odot} < M_{*} < 1.2{\times}10^6\ M_{\odot}$\ ($-9.1 > M_{V} > -10.3$). The `standard' halo occupation model typically produces more satellites than all nearby MW-mass systems, in particular producing a MW-mass galaxy with $\leqslant$\,2 satellites $<$\,0.2\% of the time. Treating each of the five known galaxies as an independent binomial trial, a $<$\,0.2\% success rate should yield one success with a $<$\,1\% probability. Moreover, there is \textit{not a single simulated galaxy} whose most massive satellite has $M_{\star}\,{\leqslant}\,10^6 M_{\odot}$. Interestingly, in our `standard' model, the typical galaxy with a satellite population similar to M94 has approximately the mass of the Large Magellanic Cloud. 

To further explore this unexpected result, we also adopt a schematic altered SMHM relation (see Figure \ref{fig:smhm}; right panel), with increased mass-dependent scatter starting at a halo mass of 10$^{11} M_{\odot}$, and a significantly steeper slope ($\sim$3) for halos $<$\,3$\times$10$^{10} M_{\odot}$, along with a 10\% probability of not forming a galaxy at all (a 10\% `failure rate'; e.g., \citealt{sawala2015}) below 10$^{10} M_{\odot}$. Figure \ref{fig:mfunc} (right panel) shows that this `stochastic' halo occupation model reproduces more accurately the typical number of satellites of a MW-mass galaxy, and gives a significantly higher likelihood of producing an M94-like system: $>$4\% of MW-mass galaxies host $\leqslant$\,2 satellites --- a $>$16\% chance for five galaxies. Additionally, several systems are produced which host a most massive satellite with $M_{\star} \lesssim 10^6 M_{\odot}$, though the probability is still $<$\,1\%. 

While this `stochastic' model is primarily used for illustrative purposes, it nonetheless strongly resembles the model used by \cite{garrison-kimmel2017} to help alleviate the TBTF problem. A broader range in the observed satellite populations around MW-mass hosts in surveys like SAGA \citep{geha2017}, and even in nearby systems excluding M94 (e.g., M101/MW vs. M81), seems to provide tentative support for this approach. The adopted slope in our stochastic model is quite similar to that of \cite{moster2013} extrapolated to lower halo masses \citep{dooley2017}. However, we find that adopting such a slope without dramatically increasing the scatter up to high masses cannot adequately reproduce M94's lack of a $M_{\star} \gtrsim 10^7 M_{\odot}$\ satellite.

To summarize, M94 directly challenges the `standard' halo occupation model. While our exploration is far from exhaustive, we find that the sparse and low-mass satellite system of M94 may indicate that galaxy formation within DM halos is much more stochastic than predicted, even for halos as massive as $\sim$10$^{11} M_{\odot}$\ --- far above the TBTF mass scale predicted to signal an increase in stochasticity by most current hydrodynamical models \citep[e.g.,][]{munshi2017,fitts2017,garrison-kimmel2018}.

\section{Conclusions}
\label{sec:con}
We have presented the discovery of two low-mass satellites of the MW-mass galaxy M94 in a deep, 150\,kpc-radius Subaru HSC survey. Both satellites have $M_{V}\,{\sim}\,{-}10$\ and $M_{\star}\,{\lesssim}\,10^6 M_{\odot}$. Both also appear to be actively star-forming, despite projected distances from M94 of $<$\,100\,kpc. 

We have conducted artificial galaxy tests and have found that our `classical' dwarf ($M_{V}\,{\gtrsim}\,{-}9.1$; $M_{\star}\,{\gtrsim}$\,4$\times$10$^5 M_{\odot}$) detection completeness is 85\% within our survey footprint up to $\sim$10$^6 M_{\odot}$\ and is $>$99\% at higher masses --- M94 very likely hosts only two `classical' satellites between projected radii of 15\,kpc and  150\,kpc. 

Furthermore, we have found that most currently accepted SMHM relations and `standard' method of DM halo occupation cannot produce a satellite population like M94's with sufficient likelihood --- $\lesssim$\,0.2\% of MW-mass central galaxies painted onto EAGLE dark matter halos host $\leqslant$\,2 `classical' satellites within 150\,kpc in projection, and \textit{none} host a most massive satellite with $M_{\star} \leqslant$\ 10$^{6} M_{\odot}$. Furthermore, `standard' halo occupation reproduces the overall satellite population of MW-mass galaxies poorly. In order to substantially increase the probability of forming an M94-like system and improve the fit to the overall population, we have presented a model which increases the scatter in the SMHM relation above 0.2\,dex for halos as massive as $10^{11} M_{\odot}$, culminating in $>$1\,dex of scatter for 10$^9 M_{\odot}$\ halos. We also increased the power-law slope of the SMHM relation to $\sim$3 for halos $<$\,3$\times$10$^{10} M_{\odot}$\ and assume that some fraction of $<$10$^{10} M_{\odot}$\ halos fail to form visible galaxies. Consequently, M94 --- a `lonely giant' which appears to only host two low-mass satellites and is completely devoid of massive companions --- may advocate for an important modification to current ideas of how the satellites around MW-mass galaxies form.\\

\acknowledgements
We thank the referee for their thoughtful feedback, which improved this paper. We also thank Annika Peter, Sergey Koposov, and Mario Mateo for helpful suggestions. AS acknowledges support for this work by the National Science Foundation Graduate Research Fellowship Program under grant No.~DGE 1256260. Any opinions, findings, and conclusions or recommendations expressed in this material are those of the author(s) and do not necessarily reflect the views of the National Science Foundation. 

Based on observations obtained at the Subaru Observatory, which is operated by the National Astronomical Observatory of Japan, via the Gemini/Subaru Time Exchange Program. We thank the Subaru support staff -- particularly Tsuyoshi Terai, Chien-Hsiu Lee, and Fumiaki Nakata -- for invaluable help preparing and carrying out the observing run. 

Based on observations utilizing Pan-STARRS1 Survey. The Pan-STARRS1 Surveys (PS1) and the PS1 public science archive have been made possible through contributions by the Institute for Astronomy, the University of Hawaii, the Pan-STARRS Project Office, the Max-Planck Society and its participating institutes, the Max Planck Institute for Astronomy, Heidelberg and the Max Planck Institute for Extraterrestrial Physics, Garching, The Johns Hopkins University, Durham University, the University of Edinburgh, the Queen's University Belfast, the Harvard-Smithsonian Center for Astrophysics, the Las Cumbres Observatory Global Telescope Network Incorporated, the National Central University of Taiwan, the Space Telescope Science Institute, the National Aeronautics and Space Administration under Grant No. NNX08AR22G issued through the Planetary Science Division of the NASA Science Mission Directorate, the National Science Foundation Grant No. AST-1238877, the University of Maryland, Eotvos Lorand University (ELTE), the Los Alamos National Laboratory, and the Gordon and Betty Moore Foundation.


\end{document}